\documentclass[aps,prd,10pt,nofootinbib,twocolumn]{revtex4-1}
\usepackage{amsmath,amssymb,amsfonts,dsfont,mathrsfs,amsthm,mathtools}
\usepackage{graphicx}
\usepackage{hyperref}
\usepackage{siunitx}
\usepackage[english]{babel}
\hypersetup{linktocpage,colorlinks=true,urlcolor=blue,linkcolor=blue,citecolor=blue}

\newcommand{\diff}[1]{\text{d}#1}
\newcommand{\rnabla}{\mathring{\nabla}}

\newcommand{\Lie}{\mathcal{L}}
\newcommand{\Lag}{\mathscr{L}}

\newcommand{\D}{\text{D}}
\newcommand*{\diag}{\operatorname{diag}}

\begin{document}

\title{Unimodular Einstein--Cartan gravity: Dynamics and conservation laws}

\author{Yuri Bonder}
\email{bonder@nucleares.unam.mx}

\author{Crist\'obal Corral}
\email{cristobal.corral@correo.nucleares.unam.mx}

\affiliation{Instituto de Ciencias Nucleares, Universidad Nacional Aut\'onoma 
de M\'exico, \\ Apartado Postal 70-543, Ciudad de M\'exico, 04510, M\'exico}

\begin{abstract}
Unimodular gravity is an interesting approach to address the cosmological constant problem, since the vacuum energy density of quantum fields does not gravitate in this framework, and the cosmological constant appears as an integration constant. These features arise as a consequence of considering a constrained volume element $4$-form that breaks the diffeomorphisms invariance down to volume preserving diffeomorphisms. In this work, the first-order formulation of unimodular gravity is presented by considering the spin density of matter fields as a source of spacetime torsion. Even though the most general matter Lagrangian allowed by the symmetries is considered, dynamical restrictions arise on their functional dependence. The field equations are obtained and the conservation laws associated with the symmetries are derived. It is found that, analogous to torsion-free unimodular gravity, the field equation for the vierbein is traceless, nevertheless, torsion is algebraically related to the spin density as in standard Einstein--Cartan theory. The particular example of massless Dirac spinors is studied, and comparisons with standard Einstein--Cartan theory are shown.
\end{abstract}

\maketitle

\section{Introduction}

Perhaps the simplest mechanism to account for the observed accelerated expansion of the Universe~\cite{Riess,Perlmutter} is to add a cosmological constant term into the Einstein--Hilbert action. However, the microscopic nature of the cosmological constant is not understood in the sense that the value obtained from the vacuum energy density of quantum fields is many orders of magnitude larger than that extracted from observations. Thus, the bare cosmological constant needs to be fine-tuned by many orders of magnitude to fit the experimental data. This is known as the cosmological constant problem (for a review see Ref.~\onlinecite{Weinberg:1988cp}). 

Unimodular gravity (UG) is an appealing proposal to tackle this problem, and it can be traced back as far as Einstein~\cite{Einstein:1919gv,Anderson:1971pn,vanderBij:1981ym,Buchmuller:1988wx,Unruh:1988in,Henneaux:1989zc,Ng:1990xz,finkelstein2001unimodular,ng2001small,ellis2011trace}. In this framework, the vacuum energy does not gravitate, and the cosmological constant merely appears as an integration constant. The main feature of UG is that it contains a nondynamical volume $4$-form, which reduces both the number of independent components of the metric and the gauge symmetries of the theory. Many interesting properties of UG theories have been studied~\cite{Alvarez:2005,Alvarez:2006,Pankaj:2012,Barvinsky:2017,Nozari:2017,Rajabi:2017}, including some of their quantum aspects~\cite{alvarez2008ultraviolet,Smolin:2009,Fiol:2010,Barcelo:2014,Padilla:2014yea,saltas2014uv,alvarez2015quantum,alvarez2015unimodular,Bufalo:2015,Eichhorn:2015,benedetti2016essential,Josset:2016vrq,percacci2017unimodular}. In addition, the unimodular constraint has been implemented in extensions of general relativity (GR), such as $f(R)$ theories~\cite{Nojiri:2015sfd}, teleparallel $f(T)$ theories~\cite{Nassur:2016yhc,Bamba:2016wjm}, and $F(\mathcal{G})$ theories~\cite{Houndjo:2017jsj}, among others.

On the other hand, gauge theories of gravity, such as Poincar\'e gauge theories~\cite{Blagojevic:2013xpa} and Chern--Simons gravity~\cite{Zanelli:2016cs}, are formulated within a geometrical structure that departs from (pseudo-)Riemannian geometry. In this geometry ---known as Riemann--Cartan---, spacetime curvature and torsion are regarded as independent quantities, and the latter is not assumed to vanish \textit{a priori}. The simplest theory within this framework is the Einstein--Cartan theory, where torsion is sourced by the nontrivial spin density of matter. In the case of Dirac spinors, when torsion is integrated out, the effective theory has a four-fermion interaction which might induce observable effects when high spin densities are involved~\cite{Hehl:1976kj}. Higher dimensional scenarios might enhance such an interaction, allowing to constrain the parameter space of such theories~\cite{Castillo-Felisola:2014iia,Castillo-Felisola:2014xba,RojasAbatte:2017vtq}. Moreover, a Peccei--Quinn mechanism~\cite{Peccei:1977hh,Peccei:1977ur} can be implemented in this formalism, to address the strong CP problem from a gravitational viewpoint~\cite{Mercuri:2009zi,Lattanzi:2009mg,Castillo-Felisola:2015ema}. Cosmological models based on Riemann--Cartan geometries have interesting features, such as avoidance of  singularities, bouncing solutions, inflationary epoch, among others~\cite{Puetzfeld:2004yg}.

The only extension to UG within Riemann--Cartan geometry has been presented in vacuum~\cite{Alvarez:2015oda}. However, the inclusion of matter with nontrivial spin density and their symmetries have not been studied. Developing such an extension, which is dubbed unimodular Einstein--Cartan (UEC), is the main goal of the present manuscript. The relevance of this study is threefold: first, many UG studies assume a conventional conservation law for matter fields, which is unjustified according to the restricted symmetries of UG. Second, the presence of spin and torsion can relax some geometrical restrictions that arise in standard UG, allowing one to search for more general solutions. Finally, the physical effects associated with UG in the presence of spin may uncover new avenues to look for torsion and vice versa. 

The manuscript is organized as follows: Sec.~\ref{formalism} presents a review of the Einstein--Cartan theory in the first-order formalism, which is the language used throughout this work. In Section~\ref{UEC}, the UEC theory is developed by considering the most general matter Lagrangian allowed by the symmetries. Sec.~\ref{Conserv} tackles the question of the matter conservation laws and their relation with the symmetries of the theory. The UEC theory coupled with Dirac spinors is studied in Sec.~\ref{example}, and the conclusions are presented in Sec.~\ref{Concl}. Finally, the main results of the paper are summarized in tensorial notation in Appendix~\ref{App}. Throughout this work, spacetime is considered to be a four-dimensional manifold $\mathcal{M}$ with trivial topology. Latin and greek characters are used for Lorentz and spacetime indices, respectively, and the set of $p$-forms is denoted by $\Omega^{p}\left(\mathcal{M}\right)$. This is the notation utilized, e.g., in Refs.~\onlinecite{Gockeler:1987an,Nakahara:2016}.

\section{First-order formalism}\label{formalism}

The first-order formalism of gravity considers two independent potentials: the vierbein $1$-form $e^a = e^{a}{}_\mu\diff{x^\mu}$, which is related to the spacetime metric $g_{\mu\nu}$ via $g_{\mu\nu} = \eta_{ab} e^{a}{}_\mu e^{b}{}_\nu$, where $\eta_{ab}=\diag\left(-,+,+,+\right)$, and the Lorentz connection $1$-form $\omega^{ab} = \omega^{ab}{}_\mu\diff{x^\mu}$, which encodes the affine structure. The vierbein and Lorentz connection transform as $1$-forms under diffeomorphisms (Diff), and as a vector and connection under local Lorentz transformations (LLT), respectively. In particular, for an infinitesimal Diff generated by the vector field $\xi$, and an infinitesimal LLT associated with $\lambda^{a}{}_b$, the transformation laws are\footnote{The LLT invariance can be enhanced to local Poincar\'e invariance by considering a nontrivial transformation of the Lorentz connection under local translations~\cite{Montesinos:2017epa}.}
\begin{align}
\label{diffcartan}
\text{Diff} &= \begin{cases}
 \delta_{\xi} e^a \;\, = \Lie_{\xi}e^a = i_{\xi}\diff{e^a} + \diff{i_{\xi}e^a } , \\ 
 \delta_{\xi} \omega^{ab} = \Lie_{\xi}\omega^{ab} = i_{\xi}\diff{\omega^{ab}} + \diff{i_{\xi}\omega^{ab}},
 \end{cases} 
 \\
 \label{LLTcartan}
\text{LLT} &= \begin{cases}
 \delta_\lambda e^{a}\;\, = \lambda^{a}{}_b e^{b}, \\
 \delta_\lambda \omega^{ab} = -\D\lambda^{ab},
 \end{cases}
\end{align}
where $\diff{}$ is the exterior derivative, $\D$ is the exterior covariant derivative with respect to the Lorentz connection, and $i_{\xi}:\Omega^{p}\left(\mathcal{M}\right)\to\Omega^{p-1}\left(\mathcal{M}\right)$ is the inner contraction such that, for $\alpha\in\Omega^{p}\left(\mathcal{M}\right)$,
\begin{align}
 i_{\xi}\alpha = \frac{1}{\left(p-1\right)!}\xi^{\mu}\alpha_{\mu\nu_2...\nu_p}\diff{x^{\nu_2}}\wedge...\wedge\diff{x^{\nu_p}}.
\end{align}

In this formalism, the curvature and torsion $2$-forms are respectively defined by Cartan's structure equations
\begin{align}
\label{curvaturedef}
 R^{ab} &=\diff{\omega^{ab}} + \omega^{a}{}_c\wedge\omega^{cb} = \frac{1}{2} R^{ab}{}_{cd}\,e^c\wedge e^d, \\
\label{torsiondef}
 T^a &= \diff{e^a} + \omega^{a}{}_b\wedge e^b \equiv \D e^a  = \frac{1}{2} T^{a}{}_{bc}\,e^b\wedge e^c.
\end{align}
The Bianchi identities take the form $\D R^{ab} = 0$ and $\D T^a = R^{a}{}_b\wedge e^b$, and, by virtue of such identities, the $\mathfrak{so}(3,1)$-valued $3$-forms
\begin{align}
\label{defmathcalea}
 \mathcal{E}_a &\equiv \epsilon_{abcd}R^{bc}\wedge e^d,\\
 \label{defmathcaleab}
 \mathcal{E}_{ab} &\equiv \epsilon_{abcd}T^c\wedge e^d, 
\end{align}
satisfy 
\begin{align}
\label{georel1}
 \D\mathcal{E}_a &= i_a R^{bc}\wedge\mathcal{E}_{bc} + i_a T^b\wedge\mathcal{E}_b,\\
\label{georel2}
 \D\mathcal{E}_{ab} &= \mathcal{E}_{[a}\wedge e_{b]},
\end{align} 
where $\epsilon_{abcd}$ is a completely antisymmetric invariant tensor such that $\epsilon_{0123}=1$, and $i_a \equiv i_{E_a}$ where $E_a = E^{\mu}{}_a \partial_\mu$ is the dual vierbein basis.

The Einstein--Cartan theory with cosmological constant belongs to the so-called Lovelock--Cartan family~\cite{Mardones:1990qc}, which describes the most general action in four dimensions such that (i)~is a polynomial on the vierbein and (derivatives of) the Lorentz connection, (ii)~is invariant under Diff and LLT, and (iii) is constructed without the Hodge dual\footnote{This restriction ensures that the field equations remain of first order, since for an arbitrary $\alpha\in\Omega^p\left(\mathcal{M}\right)$, $\diff{^2\alpha}=0$, but $\diff{\star}\diff{\alpha}\neq0$. The absence of this restriction was studied in Ref.~\onlinecite{Baekler:2011jt}} $\star:\Omega^{p}\left(\mathcal{M}\right)\to\Omega^{(4-p)}\left(\mathcal{M}\right)$, defined as acting on $\alpha\in\Omega^p\left(\mathcal{M}\right)$ as 
\begin{align}
 \star\alpha = \frac{1}{p!(4-p)!}\alpha_{a_1...a_p} \epsilon^{a_1...a_p}{}_{a_{p+1}...a_4}\,e^{a_{p+1}}\wedge...\wedge e^{a_4}.
\end{align}
The action principle for such a theory is
\begin{align}
\label{ECT}
 S &= \frac{1}{4\kappa}\int\epsilon_{abcd}\left(R^{ab} - \frac{\Lambda}{6}e^a\wedge e^b\right)\wedge e^c\wedge e^d + \int\Lag_m,
\end{align}
where $\Lag_m = \Lag_m[e^a,\omega^{ab},\phi] \in \Omega^4\left(\mathcal{M}\right)$ is the matter Lagrangian, and $\phi$ collectively denotes the matter fields.\footnote{Additional terms can be included in the action principle such as $R_{ab}\wedge e^a\wedge e^b$ and $T_a\wedge T^a$. In general, these parity-odd terms contribute to the field equations and might induce interesting effects~\cite{Hojman:1980kv,Perez:2005pm,Freidel:2005sn}. When combined as $T_a\wedge T^a - R_{ab}\wedge e^a\wedge e^b = \diff{}\left(e_a\wedge T^a\right)$ they give rise to the Nieh--Yan topological density~\cite{Nieh:1981ww}, which measures the difference between two Pontryagin classes~\cite{Chandia:1997hu}.}

The equations of motion are obtained by performing stationary variations of Eq.~\eqref{ECT} with respect to $e^a$, $\omega^{ab}$, and $\phi$, giving respectively
\begin{subequations}
\begin{align}
 \mathcal{E}_a - 2\Lambda\star e_a &= 2\kappa\tau_a,\\
 \label{cartaneq}
 \mathcal{E}_{ab} &= \kappa\sigma_{ab}, \\
 \frac{\delta\Lag_m}{\delta\phi} &= 0,
\end{align}\end{subequations}
where $\tau_a \equiv \delta\Lag_m/\delta e^a$ and $\sigma_{ab} \equiv 2\delta\Lag_m/\delta\omega^{ab}$ are the $\mathfrak{so}(3,1)$-valued energy-momentum and spin density $3$-forms, respectively. Invariance of the matter action under Diff and LLT implies the on-shell conservation laws
\begin{align}
\label{eccld}
 \D\tau_a &= i_a T^b\wedge \tau_b + \frac{1}{2}i_a R^{bc}\wedge\sigma_{bc},\\
\label{ecclllt}
 \D\sigma_{ab} &= 2\tau_{[a}\wedge e_{b]}.
\end{align}
In components, Eq.~\eqref{ecclllt} relates the spin density with the antisymmetric part of the energy-momentum tensor (see Eq.~\eqref{ecclllt_app} in the appendix). Notice that within the first-order formalism, the energy-momentum tensor is defined as the functional derivative of the matter action with respect to the vierbein, which, in principle, has no index symmetries. In the absence of spin density, torsion vanishes, the energy-momentum tensor is symmetric, and the Einstein field equations are recovered. In general, however, the spin density acts as a source of torsion, and the Einstein field equations are no longer equivalent to those derived from Einstein--Cartan theory. In the next section this theory is generalized to include the unimodular constraint.

\section{Unimodular Einstein--Cartan theory with spin density}\label{UEC}

The main feature of unimodular theories is that the metric determinant $g$ satisfies $\sqrt{-g} = \varepsilon_0$, where $\varepsilon_0$ is a nondynamical scalar density of weight $1$. Even though the original proposal was to consider this scalar density to be a constant, in this section the general case is analyzed by considering matter fields with nontrivial spin density. The unimodular constraint can be implemented through the metric redefinition~\cite{vanderBij:1981ym,Buchmuller:1988wx,Unruh:1988in,Henneaux:1989zc,Ng:1990xz}
\begin{align}
\label{metricdef}
 g_{\mu\nu} = \sqrt{\chi}\; \tilde{g}_{\mu\nu},
\end{align}
where $ \chi =\varepsilon_0/\sqrt{-\tilde{g}}$. Notice that $\tilde{g}_{\mu\nu}$ is unconstrained, and the scalar field $\chi$ measures the ratio between the nondynamical volume element and the one associated with $\tilde{g}_{\mu\nu}$. Analogously, in the first-order formalism, this field redefinition is applied at the vierbein level through
\begin{align}
\label{constr chi}
 e^a = \chi^{1/4} \tilde{e}^a,
\end{align}
where $\det e^{a}{}_\mu \equiv e = \varepsilon_0$, $\tilde{e}^a$ stands for the unconstrained vierbein, $\tilde{e}\equiv \det \tilde{e}^{a}{}_\mu$, and $\chi =\varepsilon_0/\tilde{e}$.

The main advantage of introducing $\chi$ is that it allows one to write the UEC action with the most general matter Lagrangian $\tilde{\Lag}_m$ compatible with the symmetries of the theory (see Sec.~\ref{Conserv}). This is achieved by considering $\tilde{\Lag}_m$ to have a general dependence on $\chi$. Thus, the action is
\begin{align}
\notag
 S &= \frac{1}{2\kappa}\int\bigg[\frac{1}{2}\sqrt{\chi}\epsilon_{abcd} R^{ab}\wedge \tilde{e}^c\wedge \tilde{e}^d -2\lambda \left(\chi\tilde{\epsilon} - \epsilon\right)\bigg] \\
 \label{uecaction}
 &\quad + \int\tilde{\Lag}_m[\tilde{e}^a,\chi,\omega^{ab},\phi],
\end{align}
where $\lambda$ is a Lagrange multiplier that imposes the constraint~\eqref{constr chi} on shell. Notice that the condition \eqref{constr chi} is implemented as a relation between the volume $4$-forms associated with $e^a$ and $ \tilde{e}^a$, which have the form
\begin{align}
\tilde{\epsilon} &= \frac{1}{4!}\epsilon_{abcd}\tilde{e}^a\wedge \tilde{e}^b\wedge \tilde{e}^c\wedge \tilde{e}^d,\\
\notag
 \epsilon &= \frac{1}{4!}\epsilon_{abcd} e^a\wedge e^b\wedge e^c\wedge e^d \\
 &= \frac{1}{4!}\varepsilon_0(x)\epsilon_{\mu\nu\lambda\rho}\diff{x^\mu}\wedge\diff{x^\nu}\wedge\diff{x^\lambda}\wedge\diff{x^\rho}.\label{epsilon}
\end{align}
The dependence of the Einstein--Hilbert term (i.e., the first term in the action) on $\chi$ is chosen for later convenience. In principle, however, one could consider a first-order action such that an arbitrary dependence on $\chi$ is allowed within a generic gravitational action, as expected from well-known results in scalar-tensor theories~\cite{Flanagan:2004}. In general, this class of theories can introduce interesting effects associated with the torsion~\cite{Toloza:2013wi,Espiro:2014uda,Valdivia:2017sat,Cid:2017wtf}. Even though this possibility is worth exploring, we focus on the UEC theory for the sake of simplicity, and relegate the general dependence on $\chi$ to the matter action only.

The field equations are obtained by performing stationary variations of Eq.~\eqref{uecaction} with respect to $\tilde{e}^a$, $\omega^{ab}$, $\chi$, and $\lambda$, leading respectively to
\begin{subequations}\label{generalEOMs}
\begin{align}
 \label{eome}
 \sqrt{\chi}\tilde{\mathcal{E}}_a - 2\lambda\chi\,\tilde{\star}\,\tilde{e}_a &= 2\kappa \tilde{\tau}_a,\\
 \label{eomw}
 \sqrt{\chi}\tilde{\mathcal{E}}_{ab} + \diff{\sqrt{\chi}}\wedge\tilde{\star}\left(\tilde{e}_a\wedge\tilde{e}_b \right) &= \kappa\tilde{\sigma}_{ab},\\
 \label{eomchi}
 \frac{1}{\sqrt{\chi}} \tilde{\mathcal{E}}_a\wedge \tilde{e}^a + 8\lambda\tilde{\epsilon} &= 8\kappa\frac{\delta\tilde{\Lag}_m}{\delta\chi},\\
 \label{eoml}
 \chi\epsilon &= \tilde{\epsilon},
\end{align}
\end{subequations}
where $\tilde{\tau}_a \equiv \delta\tilde{\Lag}_m/\delta\tilde{e}^a$ and $\tilde{\sigma}_{ab} \equiv 2\delta\tilde{\Lag}_m/\delta\omega^{ab}$ are the $\mathfrak{so}(3,1)$-valued energy-momentum and spin density $3$-forms, respectively. The tilde denotes quantities associated with $\tilde{e}^a$, for instance: $\tilde{\mathcal{E}}_a = \epsilon_{abcd}R^{bc}\wedge\tilde{e}^d$, $\tilde{\mathcal{E}}_{ab} = \epsilon_{abcd}\tilde{T}^c\wedge\tilde{e}^d$, $\tilde{T}^a = \diff{\tilde{e}^a} + \omega^{a}{}_b\wedge\tilde{e}^b$, and $\tilde{\star}\,\tilde{e}_a = \epsilon_{abcd}\tilde{e}^b\wedge\tilde{e}^c\wedge\tilde{e}^d/3!$, with $\tilde{\star}$ being the Hodge dual associated with $\tilde{e}^a$. Additionally, the matter field equations are assumed to hold.

As it can be noticed by inspection, not all Eqs.~\eqref{generalEOMs} are independent. For instance, $\lambda$ can be obtained by two methods: first, contracting Eq.~\eqref{eome} with $\tilde{e}^a$ and then taking the Hodge dual, which gives
\begin{align}\label{lambda1}
\sqrt{\chi}\tilde{\mathcal{E}}-8\lambda \chi = 2 \kappa \tilde{\tau},
\end{align}
where $\tilde{\mathcal{E}}=\tilde{\star}(\tilde{\mathcal{E}}_a\wedge \tilde{e}^a)$ and $\tilde{\tau}=\tilde{\star}(\tilde{\tau}_a\wedge \tilde{e}^a)$, when translated into tensor notation, are the traces of the (torsionful) Einstein and energy-momentum tensors, respectively. The second method is to take the Hodge dual on Eq.~\eqref{eomchi}, giving
\begin{align}\label{lambda2}
\sqrt{\chi}\tilde{\mathcal{E}}-8\lambda \chi = 8\kappa \chi\; \tilde{\star}\frac{\delta\tilde{\Lag}_m}{\delta\chi}.
\end{align}
The comparison of Eqs.~\eqref{lambda1} and \eqref{lambda2} yields
\begin{align}
\label{constraintmatter}
\frac{1}{4}\frac{\delta\tilde{\Lag}_m}{\delta\tilde{e}^a} \wedge\tilde{e}^a=\chi\frac{\delta\tilde{\Lag}_m}{\delta\chi},
\end{align}
which, in turn, implies
\begin{align}
\label{matterconstraint}
\tilde{\Lag}_m[\tilde{e}^a,\chi,\omega^{ab},\phi] = \tilde{\Lag}_m[\chi^{1/4}\tilde{e}^a,\omega^{ab},\phi]. 
\end{align}
Therefore, the matter Lagrangian cannot have an arbitrary dependence on $\chi$ when the field equations are satisfied. In fact, from this point on, it is assumed that Eq.~\eqref{matterconstraint} holds, which accounts for Eq.~\eqref{eomchi}. This is one of the main results of this paper and it is valid for any $\tilde{T}^a$, including the trivial case $\tilde{T}^a=0$. 

Inserting $\lambda$ in Eq.~\eqref{eome} leads to
\begin{align}
\label{tleome}
\sqrt{\chi}\left(\tilde{\mathcal{E}}_a - \frac{1}{4}\tilde{ \mathcal{E}} \tilde{\star}\tilde{ e}_a\right) = 2\kappa\left(\tilde{\tau}_a - \frac{1}{4}\tilde{\tau} \tilde{\star} \tilde{e}_a\right).
\end{align}
In terms of the constrained vierbein, which has no tilde, the UEC field equations are
\begin{subequations}\label{redefinedEOMs}
\begin{align}
\label{eome1}
\mathcal{E}_a - \frac{1}{4} \mathcal{E}\star e_a &= 2\kappa\left(\tau_a - \frac{1}{4}\tau \star e_a\right),\\
\mathcal{E}_{ab} &= \kappa\sigma_{ab},\label{TvsSigma}\\
 e &= \varepsilon_0,\label{constrainte}
\end{align}
\end{subequations}
where $\tau_a \equiv \delta\Lag_m/\delta e^a$ and $\sigma_{ab} \equiv 2\delta\Lag_m/\delta\omega^{ab}$ are defined in terms of $\Lag_m = \tilde{\Lag}_m[e^a,\omega^{ab},\phi]$. The torsion $2$-form in $\mathcal{E}_{ab}$ is such that $T^a=\D e^a$, and it is related to $\tilde{T}^a$ through
\begin{align}
 T^a = \chi^{1/4}\left(\tilde{T}^a + \diff{\ln\chi^{1/4}}\wedge\tilde{e}^a\right).
\end{align}
It has been assumed that the Lorentz connection remains unaffected by the vierbein redefinition. The UEC theory is similar to Einstein--Cartan in that the torsion does not propagate but it is algebraically related to the spin density. However, the vierbein equation is traceless and this field is constrained by Eq.~\eqref{constrainte}, which is reminiscent of the conventional UG. 

As it can be seen from Eq.~\eqref{TvsSigma}, torsion vanishes in the absence of spin density. This is not obvious since Eq.~\eqref{eomw} could suggest that $\chi$ acts as a torsion source. Therefore, in vacuum, UEC reduces to the standard vacuum UG, which is known to be equivalent to GR with a cosmological constant arising as an integration constant. This concludes the analysis of the UEC dynamics and the next section is devoted to studying the matter conservation law.

\section{Conservation laws}\label{Conserv}

Several papers in UG assume that the energy-momentum tensor is divergence free (see, e.g., Refs.~\onlinecite{Unruh:1988in,Ng:1990xz,Padilla:2014yea}). However, this is not a direct consequence of the unimodular constraint. Imposing this conservation law leads to additional restrictions that are typically disregarded. The conservation law of matter fields can be obtained by considering two approaches. The first method is to apply a covariant derivative to Eq.~\eqref{eome1} and use the Bianchi identities, leading to 
\begin{align}\label{conserveq1}
\D\tau_a -i_a T^b\wedge \tau_b -\frac{1}{2} i_a R^{bc}\wedge\sigma_{bc}=\diff{}\left(-\frac{\mathcal{E}}{8 \kappa}+\frac{\tau}{4} \right)\star e_a,
\end{align}
which implies the conservation law
\begin{align}\label{conserveq2}
 \diff{\star} i^a\left(\D\tau_a -i_a T^b\wedge \tau_b -\frac{1}{2} i_a R^{bc}\wedge\sigma_{bc}\right)=0.
\end{align}
Notice that, since this method uses the field equations, this relation must be regarded as an on-shell conservation law.

The second approach is based on the symmetries of the theory. The main advantage of this approach is that it can be used to obtain the off-shell conservation laws, which are relevant when unconventional quantum effects are considered (see for instance Ref.~\onlinecite{Josset:2016vrq}). It is clear that to apply this method is necessary to determine the symmetries of the theory. The unimodular constraint implies that only those Diff that leave the volume $4$-form $\epsilon$ invariant are symmetries of the theory.\footnote{Interestingly, breaking Diff invariance implies Lorentz violation and vice versa~\cite{KosteleckyBlumm1,KosteleckyBlumm2}. Thus, it may be possible to use experiments looking for Lorentz violation~\cite{DataTables} to put bounds on the unimodular constraint.}  Under a general Diff generated by a vector field $\xi$, such a volume $4$-form transforms as
\begin{align}\label{VPDcondition}
\delta_{\xi}\epsilon = \Lie_{\xi} \epsilon = \diff{} i_\xi \epsilon = \diff{\star}\xi.
 \end{align}
Therefore, the subgroup of Diff for which Eq.~\eqref{VPDcondition} vanishes are the symmetries of this theory. This class of Diff is known as volume preserving Diff (VPD).\footnote{Volume preserving Diff and transverse Diff are usually unequivalent, as discussed in Ref.~\onlinecite{Lopez:2011}.} In fact, to study the transformation law of the matter action under VPD, it is necessary to find a generic solution of $\diff{\star}\xi=0$. However, it is trivial to show that $\xi = \star\diff{\alpha}$, where $\alpha$ is an arbitrary $2$-form, fulfils the required condition. The matter action transformation under VPD is obtained using Eq.~\eqref{diffcartan}, provided that $\xi = \star\diff{}{\alpha}$. The result (modulo boundary terms) is
\begin{align}\notag
 \delta_{\xi} S_m &= \int \alpha\; \diff{\star i^a}\bigg( \D\tau_a - i_a T^b\wedge\tau_b \\
 \label{conssymm}
 &\qquad - \frac{1}{2}i_a R^{bc}\wedge\sigma_{bc} + \frac{\delta\Lag_m}{\delta\phi}i_a\diff{\phi} \bigg).
\end{align}
Given that $\alpha$ is arbitrary, then, on shell, Eq.~\eqref{conserveq2} is recovered.

Observe that, by virtue of the Poincar\'e lemma, there exists a $\Theta\in\Omega^0\left(\mathcal{M}\right)$ such that
\begin{equation} \label{conssymmclosed}
\star i^a \bigg( \D\tau_a - i_a T^b\wedge\tau_b - \frac{1}{2}i_a R^{bc}\wedge\sigma_{bc}  + \frac{\delta\Lag_m}{\delta\phi}i_a\diff{\phi} \bigg) = \diff{\Theta}.
\end{equation}
A comparison with Eq.~\eqref{conserveq1} reveals
\begin{align}
\label{theta}
 -\frac{\mathcal{E}}{8\kappa} + \frac{\tau}{4} = \Lambda + \Theta,
\end{align}
where $\Lambda$ is an integration constant. Therefore, the field equation for the vierbein in standard Einstein--Cartan theory is obtained, where the cosmological constant appears as an integration constant, i.e.,
\begin{align}
\label{eome2}
 \mathcal{E}_a - 2\Lambda\star e_a = 2\kappa\,\bar{\tau}_a,
\end{align}
where, as shown in Ref.~\onlinecite{percacci2017unimodular} for the torsion-free case, $\bar{\tau}_a \equiv \tau_a + \Theta\star e_a$ is an ``improved'' energy-momentum $3$-form, which is conserved according to Eq.~\eqref{eccld}. As noticed in Ref.~\onlinecite{Smolin:2009}, this equation is invariant under the simultaneous shift $\tau_a \to \tau_a + C\star e_a$ and $\Theta\to\Theta - C$, where $C$ is a constant. This can be concluded from Eq.~\eqref{eome1} as well, using the fact that the traceless part of the energy-momentum $3$-form is invariant under such a shift. This result offers an interesting perspective to address the cosmological constant problem.

Notice that, in vacuum, Eq.~\eqref{conserveq1} implies that $\diff{\mathcal{E}}=0$, which is a strong constraint on the spacetime curvature. However, since torsion and spin density are nontrivial, they allow one to relax such a constraint. Additionally, in UG and UEC, the energy-momentum tensor is generically not divergence free. This implies that, when considering homogeneous and isotropic solutions, the well-known dependences of the matter energy-density on the scale factor cannot be used, since such dependences are obtained from the divergence-free condition of the energy-momentum tensor.\footnote{We thank M. Salgado for this remark.} Given that the conservation law is nontrivial, it would be interesting to extend the method of Ref.~\onlinecite{Papapetrou} to find the equations governing the evolution of free test pointlike particles in UEC. This, however, is left for a future contribution. 

Finally, when the field equations hold, invariance under LLT implies a conservation law that coincides with Eq.~\eqref{ecclllt}. This finishes the discussion of the matter conservation laws and in the next section a concrete example where matter is described by Dirac spinors is studied.

\section{Dirac spinors in unimodular Einstein--Cartan}\label{example}

Due to their nontrivial spin density, Dirac spinors have been regarded as one of the best probes of spacetime torsion in standard Einstein--Cartan theory. Here, only massless Dirac fields are considered since, even though their mass will contribute to the energy-momentum $3$-form, it does not affect the spin density and torsion. The Dirac action for massless spinors is given by
\begin{align}
 S_\psi &= \frac{1}{2}\int\Big(\bar{\psi}\star\gamma\wedge \D\psi + \D\bar{\psi}\wedge\star\gamma\psi\Big),
\end{align}
with $\bar{\psi} = i\psi^\dagger\gamma^0$, $\gamma = \gamma_a e^a$ being the $1$-form gamma matrix such that $\{\gamma_a,\gamma_b\} = 2\eta_{ab}$, and $\D\psi = \diff{\psi} +\omega^{ab}\gamma_{ab}\psi/4$, where $\gamma_{a_1..a_p} \equiv \gamma_{[a_1}...\gamma_{a_p]}$. The Dirac equation derived from this action is
\begin{align}
 \label{dirac1}
 \star\gamma\wedge\left( \D\psi - \frac{1}{2}i_a T^a\psi\right) = 0,
\end{align}
and the energy-momentum and spin density $3$-forms are respectively given by 
\begin{align}\notag
 \tau^{(\psi)}_a &= \frac{1}{2}\bigg(\D\bar{\psi}\wedge\star\left(\gamma\wedge e_a \right)\psi - \bar{\psi}\star\left(\gamma\wedge e_a \right)\wedge \D\psi \bigg), \\
 \sigma^{(\psi)}_{ab} &= \frac{1}{2} J_5\wedge e_a\wedge e_b, 
\end{align}
where the $1$-form fermionic axial current is defined by $J_5 \equiv i\bar{\psi}\gamma_5\gamma_a\psi\, e^a$, and $\gamma_5 \equiv i\gamma_0\gamma_1\gamma_2\gamma_3$.

Torsion can be solved from the Lorentz connection field equation \eqref{TvsSigma}, giving
\begin{align}\label{torsion vs dirac spinors}
 T^a = \frac{\kappa}{2}\star\left(J_5\wedge e^a\right). 
\end{align}
Then, the effective equations where torsion is replaced by the right-hand side of Eq.~\eqref{torsion vs dirac spinors} can be found by separating the Lorentz connection into a Levi-Civita connection $\mathring{\omega}^{ab} $ plus the contorsion $ K^{ab}$, where $\mathring{\omega}^{ab}$ satisfies $\diff{e}^a + \mathring{\omega}^{a}{}_{b}\wedge e^b=0$, and $K^{ab}$ is defined by $T^a = K^{a}{}_b\wedge e^b$. The effective equations are
\begin{subequations}
\begin{align}
 \label{eomeeff}
\mathring{\mathcal{E}}_a - \frac{1}{4}\mathring{\mathcal{E}}\star e_a &= 2\kappa\bigg( \tau^{\rm(eff)}_a - \frac{1}{4}\tau^{\rm(eff)}\star e_a\bigg),\\
 \label{eomdeff}
 \star\gamma\wedge\mathring{\D}\psi &= \frac{3\kappa}{8}\,\epsilon\, \left(i^a J_5\right)\,i\gamma_5\gamma_a\psi,\\
 \label{eomleff}
 e &= \varepsilon_0,
\end{align}\end{subequations}
where the ring denotes torsion-free quantities, and 
\begin{align}\notag
\tau_a^{\rm(eff)} &= \frac{1}{2}\bigg(\mathring{\D}\bar{\psi}\wedge\star\left(\gamma\wedge e_a\right)\psi - \bar{\psi}\star\left(\gamma\wedge e_a\right)\wedge\mathring{\D}\psi \bigg) \\
& - \frac{3\kappa}{16}\,J_5\wedge\star\left(J_5\wedge e_a\right).
\end{align}
Furthermore, the effective conservation law can be written as
\begin{align}
 \star i^a \mathring{\D}\tau_a^{\rm(eff)} = \diff{\Theta},
\end{align}
where $\Theta$ can be obtained from the field equations as done in Sec.~\ref{Conserv}.

In contrast to the standard Einstein--Cartan theory, in UEC only the traceless part of the four-fermion interaction gravitates. Moreover, given a particular $\varepsilon_0$, it would be possible to implement Eq.~\eqref{eomleff} by, for example, solving for one of the vierbein components. Then, there would be additional modifications to the remaining field equations. This suggests that it may be possible to use experiments that look for torsion \cite{Yuri} to investigate if the spacetime geometry is subject to the unimodular constraint.

\section{Conclusions}\label{Concl}

In this work, the unimodular extension of Einstein--Cartan theory coupled with matter fields with nontrivial spin density is presented in the first-order formalism. This study shows that torsion can open new avenues to test the unimodular constraint and vice versa. The results can be divided into two parts: the dynamical analysis, and the conservation laws associated with the restricted symmetries of the theory.

Regarding the dynamical study, the most general dependence on the nondynamical volume $4$-form within the matter Lagrangian is considered, however, a constraint fixes its functional dependence on shell. The field equation for the vierbein turns out to be the traceless part of the standard Einstein--Cartan field equation for the vierbein, and what gravitates is the traceless energy-momentum $3$-form. The latter is insensitive to the quantum fields' vacuum energy density, offering an interesting perspective to the cosmological constant problem in the presence of spin and torsion. On the other hand, seemingly to Einstein--Cartan, torsion does not propagate since it is algebraically related to the spin density. Therefore, in the absence of spin density, this theory reduces to standard UG, which is equivalent to GR with the cosmological constant arising as an integration constant. In addition, since the unimodular constraint is implemented through a Lagrange multiplier, it shows up at the level of the field equations. However, according to the UG spirit, the nondynamical volume $4$-form should be given \textit{a priori}.

The conservation law associated with the matter energy-momentum $3$-form can be found from the field equations, or by using the symmetries of the theory, i.e., VPD. The latter method can be used in future studies where unconventional quantum effects are considered, and the (classical) field equations are not assumed to be valid at all times. Using the conservation law derived from the symmetries of the matter action, it is shown that this theory can be thought as the standard Einstein--Cartan theory, provided that the cosmological constant arises as an integration constant. The same statement holds in unimodular GR, and it is generalized to include spin density and torsion. 

As a concrete example, the case where matter fields are described by massless Dirac spinors is worked out. This case highlights the modifications that arise with respect to standard Einstein--Cartan theory, and which may generate a synergy between the unimodular constraint and torsion. Finally, combining these two extensions of general relativity may open new avenues to test the current gravity paradigm, which is certainly of great interest.

\begin{acknowledgments}
We thank B.~Ju\'arez-Aubry, T.~Koslowski, A.~Majhi, E.~Okon, M.~Salgado, and D.~Sudarsky for enlightening discussions and comments. This work is supported by UNAM-DGAPA-PAPIIT Grant No.~IA101818 and UNAM-DGAPA postdoctoral fellowship.
 \end{acknowledgments}

\appendix

\section{Summary in tensor notation}\label{App}

Here, the most important results of this work are presented in tensorial notation. First, the relation between formalisms can be obtained from the vierbein postulate 
\begin{align}
\partial_\mu e^{a}{}_\nu + \omega^{a}{}_{b\,\mu} e^{b}{}_\nu \equiv \D_\mu e^{a}{}_\nu = \Gamma^{\lambda}{}_{\mu\nu} e^{a}{}_\lambda, 
\end{align}
which, when antisymmetrizing, provides the torsion components $T^{\lambda}{}_{\mu\nu} = \Gamma^{\lambda}{}_{\mu\nu} - \Gamma^{\lambda}{}_{\nu\mu}$. The torsionful covariant derivative associated with $\Gamma$ is denoted by $\nabla$ and it acts on spacetime indices.

The Lorentz curvature given in Eq.~\eqref{curvaturedef}, in components, reads as
\begin{align}
R^{a}{}_{b\,\mu\nu} = \partial_\mu \omega^{a}{}_{b\,\nu} + \omega^{a}{}_{c\,\mu}\omega^{c}{}_{b\,\nu} - (\mu\leftrightarrow\nu), 
\end{align}
and it is related with the torsionful Riemann tensor through $R^{\lambda}{}_{\rho\mu\nu} = E^{\lambda}{}_a\, e^{b}{}_\rho\, R^{a}{}_{b\,\mu\nu}$. The torsionful Ricci tensor $R_{\mu\nu}$ and curvature scalar $R$ can be constructed in the usual way from $R^{\lambda}{}_{\rho\mu\nu}$. The Bianchi identities take the form
\begin{align}
\label{bianchicomp1}
 \nabla_{[\mu}R^{\rho\sigma}{}_{\nu\lambda]} &= T^{\gamma}{}_{[\mu\nu} R^{\rho\sigma}{}_{\lambda]\gamma}, \\
\label{bianchicomp2}
 R^{\rho}{}_{[\mu\nu\lambda]} &= \nabla_{[\mu}T^{\rho}{}_{\nu\lambda]} - T^{\gamma}{}_{[\mu\nu} T^{\rho}{}_{\lambda]\gamma},
\end{align}
which can be used to obtain
\begin{align}
\notag
 \nabla^\mu\left(R_{\mu\nu} - \frac{1}{2}g_{\mu\nu}R \right) &= T^{\lambda}{}_{\nu\rho} R^{\rho}{}_\lambda - \frac{1}{2} T^{\lambda}{}_{\rho\sigma}R^{\rho\sigma}{}_{\nu\lambda}, \\
\notag
 -2 R_{[\mu\nu]} &= \left(\nabla_\lambda + T_\lambda\right)\left(T^{\lambda}{}_{\mu\nu} + 2\delta^\lambda_{[\mu}T_{\nu]} \right),
\end{align}
where $T_\mu \equiv T^{\lambda}{}_{\mu\lambda}$ is the torsion trace. 

The field Eqs.~\eqref{redefinedEOMs} become
\begin{subequations}\label{fieldEqComponents}
\begin{align}
R_{\mu\nu}- \frac{1}{4}g_{\mu\nu}R&= \kappa\left(\tau_{\mu\nu} - \frac{1}{4}g_{\mu\nu}\tau\right),\\
T^{\lambda}{}_{\mu\nu} + 2\delta^\lambda_{[\mu}T_{\nu]} &= \kappa\sigma^{\lambda}{}_{\mu\nu},\\
 \sqrt{-g} = e &= \varepsilon_0,
\end{align}
\end{subequations}
where $\tau_{\lambda\rho}$ and $\sigma^{\mu}{}_{\lambda\rho}$ can be read from 
\begin{align}
 \tau^{\mu}{}_a \equiv \frac{1}{e} \frac{\delta\left(e\Lag_m\right)}{\delta e^{a}{}_\mu},\\
 \sigma^{\mu}{}_{ab} = \frac{2}{e}\frac{\delta\left(e\Lag_m\right)}{\delta\omega^{ab}{}_\mu},
\end{align}
and vierbein and its inverse have been used to map Lorentz indices into spacetime ones, and the spacetime metric $g_{\mu\nu}$ to lower indices. 

Finally, the conservation law associated with the invariance of the matter action under LLT is
\begin{align}\label{ecclllt_app}
 \left(\nabla_\mu + T_\mu \right)\sigma^{\mu}{}_{\lambda\rho} &= 2\,\tau_{[\lambda\rho]}.
\end{align}
The VPD are associated with vector fields that leave $\sqrt{-g}$ invariant. Since $\delta_\xi \sqrt{-g}=\sqrt{-g}\rnabla_\mu \xi^\mu $, it implies $\rnabla_\mu \xi^\mu=0$. The solution to this equation can be written in terms of a $2$-form $\alpha$ as $\xi^\mu = -\epsilon^{\mu\nu\lambda\rho}\rnabla_\nu\alpha_{\lambda\rho}/2$. Following the method outlined in Sec.~\ref{Conserv}, the conservation law associate with VPD takes the form
\begin{align}
 \left(\nabla_\mu + T_\mu\right)\tau^{\mu}{}_\rho - T^{\lambda}{}_{\rho\mu}\tau^{\mu}{}_\lambda - \frac{1}{2}R^{\lambda\nu}{}_{\rho\mu}\sigma^{\mu}{}_{\lambda\nu} = \nabla_\rho \Theta,
\end{align}
where the field Eqs.~\eqref{fieldEqComponents} and the Bianchi identities can be used to obtain
\begin{align}
-\Theta = \Lambda + \frac{1}{4\kappa}\Big( R + \kappa\, e^{a}{}_\mu \tau^{\mu}{}_a\Big).
\end{align}
This summarizes the main results of the paper in tensor notation.

\bibliography{References}

\end{document}